RESEARCH ARTICLE

# A neural field model for color perception unifying assimilation and contrast

Anna Song[1]*, Olivier Faugeras[2,3], Romain Veltz[2]

**1** Student at Département de Mathématiques et Applications, École Normale Supérieure, 45 rue d'Ulm, 75005, Paris, France, **2** MathNeuro Team, Inria Sophia Antipolis Méditerranée, 2004 Route des Lucioles-BP 93, 06902, Sophia Antipolis, France, **3** TOSCA Team, Inria Sophia Antipolis Méditerranée, 2004 Route des Lucioles-BP 93, 06902, Sophia Antipolis, France

* anna.song@ens.fr, anna.song.maths@gmail.com

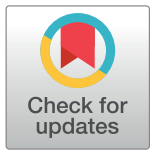

**Data Availability Statement:** The authors confirm that all data underlying the findings are fully available without restriction. All relevant files and methods are within the manuscript and its supporting information files; their data and codes are available at https://github.com/ansonang3/ colorneuralfield.git. Part of the data underlying the results presented in the study are owned by Patrick Monnier (patrick.monnier@colostate.edu) and Steven K. Shevell (shevell@uchicago.edu).

**Funding:** A.S. has received financial support from Ecole Normale Superieure of Paris as a student

## Abstract

We address the question of color-space interactions in the brain, by proposing a neural field model of color perception with spatial context for the visual area V1 of the cortex. Our framework reconciles two opposing perceptual phenomena, known as simultaneous contrast and chromatic assimilation. They have been previously shown to act synergistically, so that at some point in an image, the color seems perceptually more similar to that of adjacent neighbors, while being more dissimilar from that of remote ones. Thus, their combined effects are enhanced in the presence of a spatial pattern, and can be measured as larger shifts in color matching experiments. Our model supposes a hypercolumnar structure coding for colors in V1, and relies on the notion of color opponency introduced by Hering. The connectivity kernel of the neural field exploits the balance between attraction and repulsion in color and physical spaces, so as to reproduce the sign reversal in the influence of neighboring points. The color sensation at a point, defined from a steady state of the neural activities, is then extracted as a nonlinear percept conveyed by an assembly of neurons. It connects the cortical and perceptual levels, because we describe the search for a color match in asymmetric matching experiments as a mathematical projection on color sensations. We validate our color neural field alongside this color matching framework, by performing a multi-parameter regression to data produced by psychophysicists and ourselves. All the results show that we are able to explain the nonlinear behavior of shifts observed along one or two dimensions in color space, which cannot be done using a simple linear model.

## Author summary

The color perception produced by an image heavily depends on the spatial distribution of its colors. From this "color in context" phenomenon, extensively studied in psychophysics for decades, has arisen the question in neuroscience of how color and space interact in the brain. Visual signals are indeed processed in such a way that neighboring pixels make the perception at some point different from its real color, inducing a color shift. In this work, we propose to emulate perception in context by modeling the activity of color sensitive neurons with a neural field. Our framework unifies two antagonistic effects, assimilation civil servant. O.F and R.V. have received funding from the European Union's Horizon 2020 Framework Programme for Research and Innovation under the Specific Grant Agreement No. 785907 (Human Brain Project SGA2). The funders had no role in study design, data collection and analysis, decision to publish, or preparation of the manuscript.

**Competing interests:** The authors have declared that no competing interests exist.

and contrast, which have been suggested to occur simultaneously but at different scales. We use the notion of color opponency inspired by the work of Hering, so as to express these effects as a combination of attraction and repulsion in physical and color spaces. We introduce the concept of "color sensation", and show how to rigorously link the neural field model to perceptual shifts, by considering color matching as a mathematical projection on color sensations. The results show that our model is able to reproduce some non-trivial behaviors of the color shifts observed in experiments.

## Introduction

Color induction, which refers to a change in color appearance of a test stimulus under the influence of spatially neighboring stimuli in the field of view [1], has been extensively studied in psychophysics [2]. This effect has been observed for uniform inducing surrounds [3–7] and geometrically more complex ones as well [8–15]. The geometry of spatial context, and especially the frequency of chromatic modulation, have been shown to play an important role in color induction. Many works on this subject have been devoted to the study of two induction effects in particular, known as *chromatic assimilation* and *simultaneous contrast* (see Fig 1). Chromatic assimilation is the fact that the chromatic appearance of a test stimulus changes *towards* the chromaticity of inducing stimuli. Conversely, simultaneous contrast corresponds to the test chromatic appearance changing *away from* the chromaticity of inducing stimuli. Contrast is interesting in that it involves the notion of *color opponency*: the change is often made towards an opponent or complementary color [5, 7, 16]. These increased perceptual similarity or dissimilarity can be viewed as the results of attractive or repulsive effects respectively in color space.

While assimilation and contrast had previously been thought to occur separately, [15, 17, 18] suggested that they act simultaneously in a synergistic manner. The idea that effects induced by context result from a balance between assimilation and contrast was also proposed in cognition [19]. The experimental settings used by [18] to demonstrate the synergy relied on *asymmetric color matching*. In psychophysics, this is a classical procedure to objectively measure the amount of color induction caused by spatial context. A human observer views two still color images side by side, a *test image* $I^{test}$ whose pattern influences the perception of a test color $c^{test}$, and a *comparison image* $I^{comp}[c^{comp}]$ with a modifiable comparison color $c^{comp}$ (see Fig 2). In most experiments, the two images have the same geometric patterns and are composed of elementary shapes, such as rectangular, round or concentric patches, uniformly filled with different colors [3, 7, 20]. The patches to be compared are filled with $c^{test}$ and $c^{comp}$. The observer is asked to change $c^{comp}$ until color appearance between the test and comparison patches are the same, leading to a *perceptual match*. The perceptual *shift* is then the difference between the final color $c^{match}$ and the test color $c^{test}$. In Fig 3, we illustrate this experiment with simple square patterns.

In [18], color shifts were measured for patterns with concentric annuli as in Fig 1 (Down), whose colors were distinguished by the stimulation of $S$ cones only. The matching surround was a neutral gray. Stimuli were expressed in a variant of the cone-based chromaticity space proposed by [22] and specified by three coordinates $s, l, Y$, where $s$ and $l$ are defined as the ratios $\frac{S}{L+M}$ and $\frac{L}{L+M}$ respectively, and $Y$ stands for luminance. The definition of the $S, M, L$ signals is recalled below. Their results showed that the largest color shifts in $s$ chromaticity were induced by patterns alternating between two distinct colors, such as purple and lime. In particular, shifts induced by uniform backgrounds or patterns alternating between white and purple or lime were smaller.

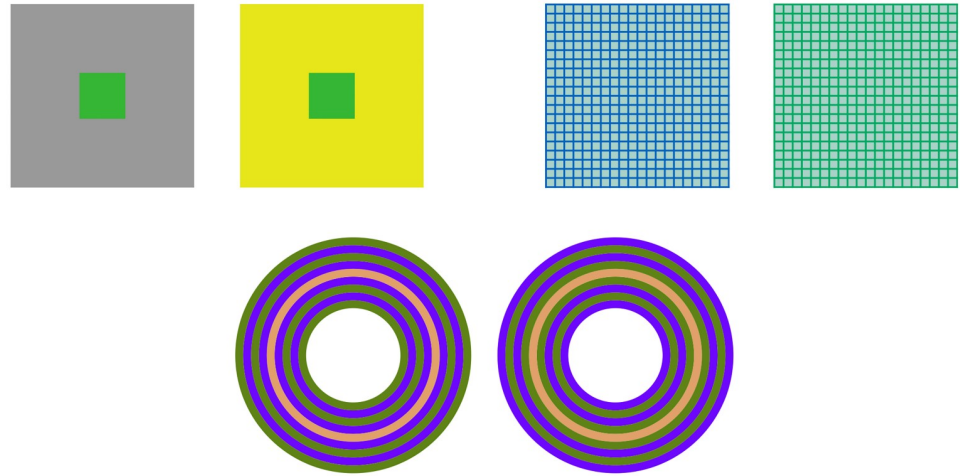

**Fig 1. Assimilation and contrast in action (in the style of [17] Fig 1).** Up, left. Simultaneous contrast: the small patches on the left and the right are *identical*, but they tend to *appear* darker inside the yellow surround, lighter inside the dark surround. Up, right. Chromatic assimilation: the same grayish background tends to be blue or green depending on the color and spatial frequency of the grid covering it. Down. Synergy of both phenomena: the two central rings are identical, but are perceived as pink or orange when surrounded by concentric annuli with a purple/lime or lime/purple pattern.

Such large shifts could not be induced by optical factors (spread light or chromatic aberration) only, but implied some neural processing of the stimuli [23]. To explain this, [18] suggested that $s^{test}$ was shifted towards the adjacent ring thanks to assimilation, while it was also repelled away from the second ring by contrast, resulting in the matching value $s^{match}$. They proposed a S+/S- center-surround receptive field model to predict the shifts: at a point $x$ in the test ring,

$$\text{shift at } x := s^{match} - s^{test} = \text{DOG} * (J^{test} - J^{comp}[s^{test}])(x),$$

where $J^{test}$ and $J^{comp}$, in $s$ coordinates, are convolved with a Gaussian kernel beforehand to account for retinal blurring; the Difference of Gaussians DOG stands for the receptive field. Color appearance would then be the combined contributions of chromaticity at the point of interest and influence of the surround filtered by the S+/S- receptive field.

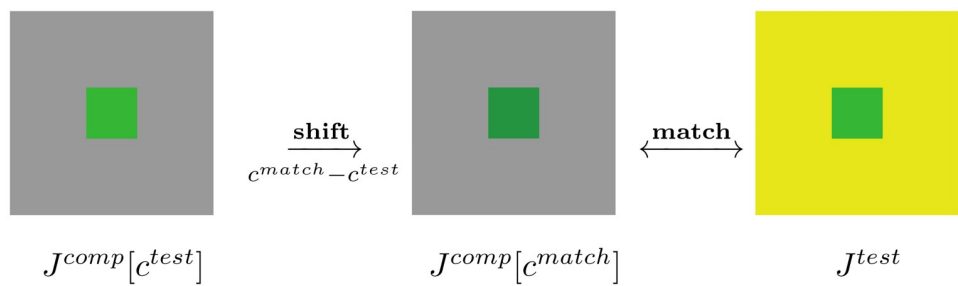

$$J^{comp}[c^{test}] \qquad J^{comp}[c^{match}] \qquad J^{test}$$

**Fig 2. Color matching measures the influence of context over color perception.** The green color $c^{test}$ is tested against a yellow surround in the test image $J^{test}$. The observer changes the comparison color $c^{comp}$ inside $J^{comp}$ until a match occurs at $c^{match}$ between the two central patches. Color induction is measured as the obtained shift.

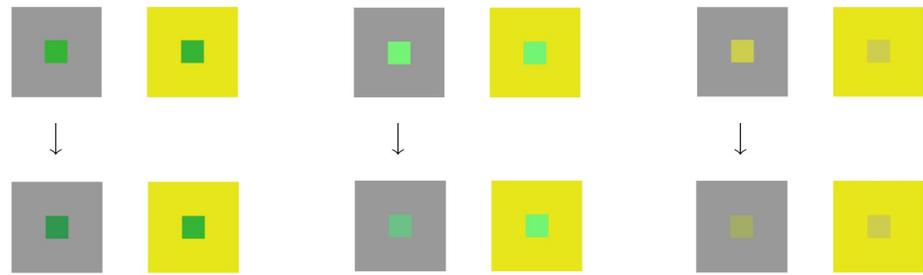

**Fig 3. A typical color matching experiment.** Upper row, left to right. For each of the three pairs of (gray, yellow) large squares, the central patches have identical colors. Their *HSL* coordinates are (120°, 55%, 46%), (120°, 85%, 71%), (60°, 57%, 56%), respectively. The HSL color space is extensively used in computer graphics and defined in [21]. Lower row, left to right. One of the authors (A.S.) has modified the *HSL* coordinates of the left central patch, in order to obtain perceptually equal patches (as much as possible). This is called *matching*. We can see how far the "perceived" *HSL* are from reality: (138°, 55%, 39%), (140°, 41%, 59%), (166°, 29%, 54%). The reader probably does not perceive the two patches as perceptually equal, because matching is subject-dependent.

After tuning the parameters, this simple linear model was capable of explaining data in the specific experimental setting of [18], where the chromaticity $s$ of the test ring was fixed while that of the surround was changed. However, it was unable to explain the dependency of shifts upon $s^{test}$ reported in a later paper [17] where, conversely, $s^{test}$ was varied while the surround was left unchanged. Indeed, in the above equality, the central ring of the difference image $J^{test} - J^{comp}[s^{test}]$ has chromaticity $s = 0$, after cancellation of the test and comparison chromaticities. The model is also conceptually difficult to justify, because it treats the central ring and the surround as fundamentally different spatial components, while handling single chromaticities and spatial integrations at the same conceptual level. This prevents their computational framework from being extended to various patterns and other matching experiments.

Here, we build a framework dedicated to general color matching experiments. It is able, in particular, to explain the nonlinear behavior of color shifts found by [17, 18]. As a starting point, we reformulate their fundamental observation into a **Principle of Synergy** which relies on the notion of color opponency introduced by Hering [16]:

1. Adjacent neighbors surrounding a spatial point $x$ in an image $I$ tend to perceptually *attract* towards their color, in the sense that they contribute to make the color appearance at $x$ more similar to theirs.

2. Remote neighbors tend to *repel* towards their respective *opponent* color. They are not immediately adjacent to $x$, but at some short distance.

3. Far neighbors are too far from $x$ to have any substantial influence on the color perception at $x$.

This viewpoint implies an appropriate *change of vocabulary*: in the sequel, chromatic assimilation and simultaneous contrast refer to local interactions, which may act at the same time but at two different local scales. The global effect observed is then the integration of infinitesimal influences induced by spatially neighboring points. Perception can in particular result in attraction (assimilation wins over contrast), repulsion (contrast wins over assimilation), or none of them. Thus assimilation and contrast, which seem to be contradictory effects, can be described as concomitant local phenomena.

*Beyond giving a merely computational model, we aim at designing a framework consistent with the physiological and anatomical observations currently available.* Light entering the eye stimulates L, M and S cones of the retina proportionally to quantal absorption rate [24], in a

"wavelength-blind" fashion according to the principle of univariance [25]. At a point $\mathbf{x}$ of the retina $\mathbb{R}$, the stimulus $L^{\mathbf{x}}$ of L cones can be approximated at first order as [26, 27]

$$L^{\mathbf{x}} := \int_{\lambda \in \Lambda} \mathcal{C}(\lambda) \mathcal{S}_L^{\mathbf{x}}(\lambda) \, d\lambda = \langle \mathcal{C}, \mathcal{S}_L^{\mathbf{x}} \rangle_{\mathrm{L}^2(\Lambda)}, \tag{1}$$

where $\mathcal{C}$ is the spectral distribution of the light over the visible spectrum $\Lambda$, and the spectral sensitivity $\mathcal{S}_L^{\mathbf{x}}$ of L cones located at $x$ depends on their local density.

$L$, $M$, $S$ signals are then relayed by the Lateral Geniculate Nucleus (LGN) and transmitted to the primary visual cortex (V1) through axonal projections. It has been established that chromatic input to the visual cortex from the LGN is encoded in an opponent fashion as Hering postulated [16] and as was later confirmed by neurophysiologists [28–30]. This justifies the use of a color-opponent framework here. LGN cells have been found to respond to *linear* combinations of $L$, $M$, $S$ stimuli [23, 30]. $L - M$ and $S - (L + M)$ signals are transmitted to *single-opponent* cells in layers ($4C\beta$) and ($2/3$ and $4A$) of the visual cortex respectively [2, 23]. In contrast to them, cells clustered inside or around Cytochrome Oxydase (CO) blobs in layer 2/3 of V1 are sensitive to a *continuum* of colors instead of three cardinal axes, *nonlinearly* with respect to cones [23, 31, 32]. As such, they may have a prominent function in encoding color in the cortex, as proposed in [33, 34]. Most of them were found to be *double-opponent cells* [2, 35, 36]. Single- and double-opponent cells are very likely to play a fundamental role in color processing in the brain [2, 23]. The former are important for analyzing color in large areas while the latter are sensitive to edges and orientation, implying that color assimilation would be due to single-opponent cells and color contrast to double-opponent cells [2].

The visual cortex has the specificity to be organised into *hypercolumns*, *i.e.*, groups of neurons sharing the same receptive field and coding for a particular physical quantity at this position, such as orientation, spatial frequency, and temporal frequency [32, 37–39]. These signals are mapped from the retina to V1 following an approximately logarithmic retinotopy [40, 41]. Unlike in the case of orientation, for which the existence of such hypercolumns in V1 is now well established [32], the anatomical and physiological bases for a functional architecture encoding color are still debated. However, in light of the promising findings made by [31], and as discussed in [23, 32, 42], it is reasonable to assume in our work a **hypercolumnar organisation of cells tuned to a continuum of colors**, having double-opponent characteristics, and related to CO blobs in layer 2/3 of V1. Our work also supposes the presence of long-range lateral connections between hypercolumns, in agreement with observations of [43] where horizontal connections tend to link blobs to blobs. Note that we do not use further assumptions about the anatomical organization of color-tuned cells with respect to blobs (inside, around, independent), which is still unclear [2, 23].

In this context, our model relies on a neural field [44–47]. It is worth noting that neural fields have been previously applied to simpler examples of sensory processing in visual cortex, in order to study the spontaneous formation of population tuning curves. Orientation tuning has been addressed in an important paper by Ben-Yishai and colleagues [48]. Their model of a single cortical hypercolumn did not take into account the spatial relations between these hypercolumns, which was done in the 2001 landmark paper by Bressloff and colleagues [49]. The problem was then revisited by Bressloff [50–52], and later by two of the authors of the present paper [53]. Spatial frequency tuning has been addressed by Bressloff and Cowan [54, 55], and by Chossat and Faugeras [56]. The combination of orientation tuning in binocular vision giving rise to rivalry waves has been studied in [57]. These simpler models have the benefit of explicit knowledge regarding feature preference maps (image orientations, image textures) and connectivity in V1 (orientation hypercolumns and their relations with CO blobs).

This color neural field framework allows us to study color-space interactions. In recent years, the interactions between color, and orientation/form/space, have received increasing attention [2]. Double-opponent cells may strongly contribute to the relationship between color and form processing, since the shape of their receptive fields determines their orientation tuning [23]. This supports the hypothesis that functional architectures for color and orientation would be intermingled and realize color-form interactions. A framework was proposed by [42] for modeling color and orientation processing in V1. They assume that two populations of neural masses, one color-insensitive but orientation-tuned, and the other sensitive to both, interact through an extended version of the ring model [48]. Our work is related but complementary to theirs, because rather than considering only two hypercolumns of each kind and using uniform inputs, without introducing space, we study the interactions between color and space by the means of multiple color hypercolumns and patterned inputs, without introducing orientation. The relation between orientation and color falls outside the scope of this work, and should be examined in the future.

*The goal of our work is therefore to provide a neural field model unifying assimilation and contrast inside a color-opponent framework, consistent with psychophysical data, and compatible with the physiology of V1. We have successfully achieved this aim.*

## Materials and methods

### Ethics statement

A.S. conceived and designed the experiments herself and consented to participate.

### Color and opponent representation

The rigorous definition of color is thoroughly explained in S1 Appendix, where we also present the most common representations of the color space. Here, for the sake of simplicity, we only briefly state the minimal definitions and properties to be used in the model.

*Color* is mathematically defined as an equivalence class of metameric lights [26, 58–60]. Two physical lights of spectral distributions $\mathcal{C}_1, \mathcal{C}_2 \in \mathbb{L}^2(\Lambda)$ are *metameric* if they produce exactly the same visual effect under the same viewing conditions, and this identification strongly depends on the observer. In our framework, metamerism can be expressed as the equality of the triplets of scalar products characterizing $\mathcal{C}_1$ and $\mathcal{C}_2$

$$(L, M, S) = (\langle \mathcal{C}_i, \mathcal{S}_L \rangle_{\mathbb{L}^2(\Lambda)}, \langle \mathcal{C}_i, \mathcal{S}_M \rangle_{\mathbb{L}^2(\Lambda)}, \langle \mathcal{C}_i, \mathcal{S}_S \rangle_{\mathbb{L}^2(\Lambda)}),$$

with $\mathcal{S}_L$ the spectral sensitivity of L cones (and likewise for M and S cones), see (1). We dropped the exponent $\mathsf{x} \in \mathbb{R}^2$ by considering for simplicity that cone density is constant across the retina and that light is spatially uniform (although we can define metamerism with respect to any $\mathsf{x}$). Color is hence naturally identified to a three-dimensional vector for trichromats, being specified by a triplet of cone stimuli $(L, M, S)$. For a light of spectral distribution $\mathcal{C}$, its color is denoted $[\mathcal{C}]$.

*Color space*, denoted $\mathfrak{C}$, is then defined as the subset of physically realizable colors which are visible to the eye. Since the cone signals $(L, M, S)$ are non-negative and cannot reach all possible positive values [60], $\mathfrak{C}$ can be identified to a bounded subset of $(\mathbb{R}^+)^3$ through a choice $\phi_{LMS} : \mathfrak{C} \to \mathbb{R}^3$ of coordinates. In this work, we suppose that an appropriate choice of coordinate system $\phi_{opp} : \mathfrak{C} \to \mathbb{R}^3$ leads to an *opponent representation* of the color space

$$\mathfrak{C}_{opp} := \phi_{opp}(\mathfrak{C}) \tag{2}$$

satisfying the following important properties. First, it is a bounded and convex subset of $\mathbb{R}^3$

which enjoys symmetry: if $c \in \mathfrak{C}_{opp}$, then $-c \in \mathfrak{C}_{opp}$. Second, the symmetry operation $c \mapsto -c$ must pair any color to its opponent one, in the sense of Hering [16]. Hence, color regions of $\mathfrak{C}_{opp}$ come into opposed pairs, for example Yellow and Blue or Red and Green regions, or likewise. Third, $\mathfrak{C}_{opp}$ must contain the neutral or zero color 0, opponent to itself, which would correspond to some neutral gray with no hue (for a fourth condition, see details in S1 Appendix).

Following up on this, we consider in this work an opponent representation in the style of Hering's theory. Indeed, in view of the physiological results exposed in the Introduction, it is now accepted that $L$, $M$, $S$ signals are recombined in area V1 into *Yellow-Blue*, *Red-Green* and *Achromatic* independent channels [61], although the right choice of the opposite axes has been debated [62]. Our model does not depend in a decisive manner upon the choice of a specific color opponent space. In fact our particular choice of Hering's coordinates is not very different from the cone opponency coordinates defined in [30]. Also, we do not require that the opponent axes point towards perceptually unique hues, unlike in the original theory of Hering. Supposing such a simple relationship between physiology of the cells and psychology related to hue pureness was indeed criticized [23, 30, 62].

Here, we rely on the $(l, s, Y)$ and $(H, S, L)$ representations, and restrict ourselves to a *lower-dimensional* subspace of the original color opponent space, also denoted $\mathfrak{C}_{opp}$ for convenience (in this context, the letters 'S' and 'L' of HSL stand for Saturation and Luminance respectively, not Short or Long cones, while 'H' stands for Hue).

In the case of the $(l, s, Y)$ representation, we apply our model to the *one-dimensional* subspace based on the chromaticity $s$ and defined by $c := s - 1$. The $(l, s, Y)$ representation used in [15, 17, 18], a variant of the one proposed by [22], is defined as

$$\begin{cases} s &= \frac{S}{L+M} \\ l &= \frac{L}{L+M} \\ Y &= L+M+S \end{cases}.$$

We then define $\mathfrak{C}_{opp}$ to be the one-dimensional color subspace based on the change of coordinates $c := s - 1 \in \mathfrak{C}_{opp} := [-2, 2]$, where the number 2 is arbitrary, but covers the typical range of $c$ values used in experiments (purple, lime and white correspond to $(l, s, Y) = (0.66, 2.0, 15cd/m^2)$, $(0.66, 0.16, 15cd/m^2)$ and $(0.66, 0.98, 15cd/m^2)$ respectively.).

The Hue, Saturation and Luminance or $(H, S, L)$ representation (note that the letters 'S' and 'L' are not referring to Short or Long cones), often used in computer graphics, maps the *sRGB* unit cube or *gamut* of a device to a cylinder whose central axis is achromatic and perpendicular to a chromatic disk [58, 63]. Standard formulas provide the change of coordinates $\mathcal{T}_{sRGB \to HSL}$ [21].

We use the two-dimensional chromatic disk $\mathfrak{C}_{opp}$, defined as the intersection of the constant luminance plane $L = 1/2$ with the HSL cylinder, and identified to the unit disk, so that $(c_1, c_2) := (S \cos(H), S \sin(H)) \in \mathfrak{C}_{opp}$. The gamut corresponds in fact to a subject- and device-dependent subspace $\mathfrak{C}_{dev}$ strictly smaller than the subspace of chromatic colors visible by the observer, since they are not all reproducible by a screen. The gamut of standard devices however covers a large part of visible colors, hence justifying its use, and the HSL representation has already proven its efficiency in computer graphics. *We claim that the specific details of the display, such as gamut and screen characteristics, do not play a major role in our methods and results*, provided that all experiments are consistently made in the same conditions.

**Table 1. Mathematical notations.**

| Physical space | |
|---|---|
| $\mathsf{R} \subset \mathbb{R}^2$ | spatial domain of the retina |
| $\Omega \subset \mathbb{R}^2$ | spatial domain of color hypercolumns |
| $\mathsf{x} \in \mathsf{R}$ | a retinal point |
| $r \in \Omega$ | a cortical point or hypercolumn |
| **Color space** | |
| $\mathcal{C}$ | spectral distribution of a light |
| $\mathfrak{C}$ | color space: the set of human-visible colors |
| $c \in \mathfrak{C}$ | a color |
| $\mathfrak{C}_{LMS} \subset \mathbb{R}^3$ | LMS standard representation |
| $\mathfrak{C}_{opp} \subset \mathbb{R}^3$ | an opponent representation of $\mathfrak{C}$ |
| $(L^{\mathsf{x}}, M^{\mathsf{x}}, S^{\mathsf{x}}) \in \mathfrak{C}_{LMS}$ | L,M,S signals received at $\mathsf{x} \in \mathsf{R}$ |
| **Dynamic entities** | |
| $J(\mathsf{x}, t)$ | retinal image |
| $I(r, t)$ | cortical image |
| $a(r, c, t)$ | neural activity of neural mass $(r, c)$ at time $t$ |

## Model

The main thrust of our model is *twofold*: first, we put forward an evolution equation for the dynamics of neural activities, considered as a *spatial and color neural field* [44–47]. Second, we propose a *theoretical framework for color matching experiments* in the context of color perception, and introduce a *formal definition of color sensation*.

**Notations.** The retina $\mathsf{R} \subset \mathbb{R}^2$ maps onto the spatial domain of the cortex $\Omega \subset \mathbb{R}^2$ through axonal projections. The scene projects onto the retina as a retinal image $J : \mathsf{R} \to \mathfrak{C}_{opp}$, and is transmitted to the cortex as a cortical image $I : \Omega \to \mathfrak{C}_{opp}$. Both images $I$ and $J$ take values in the opponent representation, and can be considered as triplets of scalar images. In the following, $c \in \mathfrak{C}_{opp} \subset \mathbb{R}^d$ implicitly refers to the associated color $[\mathcal{C}] = \phi_{opp}^{-1}(c)$, where $\mathfrak{C}_{opp}$ is defined in Eq (2) and $d$ depends on the dimension of the color space being considered, *i.e.* 1 or 2.

Under the assumption of color-coding hypercolumns, as discussed in the Introduction, $(r, c)$ denotes the *neural mass* selective for cortical position $r \in \Omega$ and color $c \in \mathfrak{C}_{opp}$. We define a neural activity $a$ which depends on position $r \in \Omega$, color $c \in \mathfrak{C}_{opp}$, and time $t \in \mathbb{R}$. A summary of notations used throughout the paper is given in Table 1 above.

**Color neural field.** We now describe our neural field model. We assume that, on a time interval $\mathfrak{I} \subset \mathbb{R}$ containing 0, the neural activity $a : \Omega \times \mathfrak{C}_{opp} \times \mathfrak{I} \to [0, 1]$ is solution to an integrodifferential equation of Wilson-Cowan type [64]:

$$\tau \frac{da}{dt} = -a(t) + F(\omega \star a(t) + H) \qquad a(t) \in \mathbb{L}^\infty(\Omega \times \mathfrak{C}_{opp}) \qquad (3)$$

where at each instant $a(t)$ takes values in $[0, 1]$ and represents a firing rate, or any physical activity.

- The typical speed of the dynamics $\tau$ is here of less importance than other parameters, so that it can be taken as $\tau = 1$ up to rescaling of the time axis;

- the *activation function* $F$ is a sigmoid converging to 0 and 1 at $\pm\infty$:

$$F(x) := \frac{1}{1 + e^{-\gamma x}}, \qquad (4)$$

parameterized by $\gamma$, which is proportional to the highest slope of the sigmoid $F'(0) = \frac{\gamma}{4}$;

- $H$ stands for the *color input* relayed by single-opponent cells in the LGN (alternatively, by those in layers (4$C\beta$) and (2/3 and 4$A$), see Introduction) to the neural masses $(r, c)$

$$H(r, c, t) := h(c - I(r, t)), \qquad h(c) := \mu_h e^{-\frac{\|c\|^2}{2\sigma_h^2}}, \qquad (5)$$

where the cortical image $I(r, t)$ is in opponent coordinates. Thus, the strongest input is given to neural masses sensitive to colors closest to the actual viewed color $I(r, t)$, as expected for color-tuned cells. The euclidean metric of $\mathbb{R}^3 \supset \mathfrak{C}_{opp}$ serves to compare them through $h$.

The connectivity kernel $\omega$ is the central part of our model and is designed so as to *encode the antagonistic actions* of contrast and assimilation. It acts on $a$ according to

$$\omega \star a(t) = \int_\Omega \int_{\mathfrak{C}_{opp}} g(r - r') f(c, c') a(r', c', t) \, dr' dc', \qquad (6)$$

where the $\star$ operation depends on the opponent representation $\mathfrak{C}_{opp}$, and the different functions are such that (see Fig 4)

- $g$ is a classical difference of gaussians or "Mexican hat", parameterized by weights $\mu, \nu$ and variances $\alpha, \beta$:

$$g(r) := \mu e^{-\frac{\|r\|^2}{2\alpha^2}} - \nu e^{-\frac{\|r\|^2}{2\beta^2}}. \qquad (7)$$

To have local excitation, we suppose $g(0) > 0$, *i.e.*, $\mu > \nu$. The kernel $g$ weights the *influence of spatially neighboring hypercolumns*;

- $f(c, c')$ is a function of two variables in $\mathfrak{C}_{opp}$, parameterized by $\mu_c, \nu_c, \alpha_c, \beta_c$:

$$f(c, c') := \mu_c e^{-\frac{\|c - c'\|^2}{2\alpha_c^2}} - \nu_c e^{-\frac{\|c + c'\|^2}{2\beta_c^2}}. \qquad (8)$$

Formulated as such, $f$ is symmetric in $c$ and $c'$. For any fixed $c$, $f(c, \cdot)$ is a difference of gaussians, one which is centered at $c$, the other one at its opponent $-c$. By introducing the gaussian kernels

$$f_1(c) := \mu_c e^{-\frac{\|c\|^2}{2\alpha_c^2}} \qquad f_2(c) := \nu_c e^{-\frac{\|c\|^2}{2\beta_c^2}},$$

it reformulates as

$$f(c, c') := f_1(c - c') - f_2(c + c') = f_1(c' - c) - f_2(c' - (-c)).$$

$f(c, c')$ hence measures the influence of $c'$ over $c$, depending on the position of $c'$ relative to the opponent pair $(c, -c)$. Here, the minus sign in $-c$ is essential and expresses color opponency.

Given these definitions and for $I$ regular enough, a solution to (3) exists on $\mathbb{R}$ and is unique. Because $-a < \dot{a} < 1 - a$, an elementary proof shows that it remains bounded by 0 and 1,

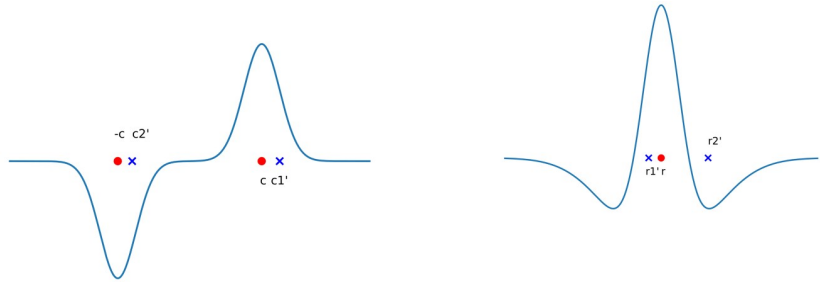

The function $f(c, \cdot)$ for fixed $c$. $c'_1$ and $c'_2$ are resp. close to $c$ or its opponent $-c'$.

The function $g(r - \cdot)$ for fixed $r$. $r'_1$ and $r'_2$ are resp. adjacent and remote neighbors.

**Fig 4. $f$ and $g$ functions, displayed on a 1D axis for illustration purpose.** The influence of $c'$ and $r'$ over $c$ and $r$, respectively, depends on their position relative to $(c, -c)$ and $r$. In color space, it is positive when $c'$ is closer to $c$, negative when it is closer to $-c$; in physical space, it is positive when $r'$ is an adjacent neighbor of $r$, and negative when it is a remote one.

provided that $0 < a(0) < 1$. In the sequel we write

$$\omega(r, c, r', c') \; := \; g(r - r') f(c, c')$$

for the connectivity kernel. The neural field dynamics (3) can then be reformulated as

$$\frac{da}{dt}(r, c, t) \; = \; -a(r, c, t) \; + \; F\left( \iint_{\Omega \times \mathscr{C}_{opp}} \omega(r, c, r', c') a(r', c', t) \, dr' dc' + H(r, c, t) \right).$$

Because we only consider static and not dynamic images here, the color input $H(r, c, t) = H(r, c)$ is kept constant.

**Interpretation of the connectivity kernel.** The connectivity kernel $\omega$ is the most important item in the neural field model, because it both expresses the antagonism between contrast and assimilation, and reflects properties reminiscent of center-surround double-opponent cells. The lateral connection from $(r', c')$ to $(r, c)$ is excitatory or inhibitory if $\omega(r, c, r', c')$ is positive or negative, respectively, and with a strength proportional to the absolute value. The level of activity $a(r', c', t) > 0$ of neighboring masses $(r', c')$ is weighted by $\omega(r, c, r', c')$, whose sign depends on the relative positions of $r'$ and $r$, $c'$ and $\pm c$. Four situations are possible, according to the respective signs of $g(r - r')$ and $f(c, c')$, summarized in Table 2.

Hence, configurations in which $(r', c')$ excites $(r, c)$ correspond to situations where $r'$ is adjacent to $r$, i.e., $g(r - r') > 0$, and $c'$ is close to $c$; or where $r'$ is a remote neighbor, i.e., $g(r - r') < 0$, and $c'$ is close to the opponent color $-c$ (see Fig 4). Otherwise, the connection is inhibitory. This behavior typically models the synergy principle of assimilation and contrast. Interestingly, this seems to suggest a behavior compatible with that of double-opponent cells. Qualitatively, the activity of $(r, c)$ is likely to increase or decrease for a center-surround ON/OFF pattern, while reaching less extreme values for uniform inputs because of compensation. In the Results section, we illustrate the roles of the connectivity kernel $\omega$ and the cortical input $H$.

**Table 2. Sign of the connectivity kernel $\omega$.**

| $g(r - r') f(c, c')$ | $c'$ close to $c$ | $c'$ close to $-c$ |
|---|---|---|
| $r'$ close to $r$ | >0 | <0 |
| $r'$ far from $r$ | <0 | >0 |

**Convolution form.** Using that

$$\omega(r, c, r', c') = g(r - r')[f_1(c - c') - f_2(-c - c')],$$

the double integral can be rewritten into a convolution-like form:

$$\omega \star a = g \underset{\Omega}{*} \left[ f_1 \underset{\mathfrak{C}_{opp}}{*} a - \mathrm{Sym} \cdot [f_2 \underset{\mathfrak{C}_{opp}}{*} a] \right],$$

where convolutions are computed in their respective spaces, and the symmetry in color space $\mathrm{Sym} = -Id_{\mathfrak{C}}$ induces a representation $(\mathrm{Sym} \cdot a)(r, c) \coloneqq a(r, -c)$ on the space of cortical activities. Note that $\mathrm{Sym} \cdot [f_2 \underset{\mathfrak{C}_{opp}}{*} a] = f_2 \underset{\mathfrak{C}_{opp}}{*} (\mathrm{Sym} \cdot a)$ thanks to symmetry of $f_2$ and $\mathfrak{C}_{opp}$.

## Color sensation and color matching experiments

The Color Neural Field Eq (3) describing (at least, theoretically) how the visual cortex reacts to a color image, we now link our model to psychophysical data. This subsection introduces the central notion produced by our model, *i.e.*, that of color sensation. Basically, it corresponds to some feeling produced in the brain when observing a still image. We define it to be a steady state of the neural field dynamics of Eq (3), then restricted to the hypercolumn in correspondance to the test point. This concept allows us to propose a mathematical description of color matching experiments (see Introduction), where matching is considered as the projection of the test color sensation onto a family of color sensations elicited by comparison images. Such a "matching as a projection" framework allows the model to predict color shifts. It could be generally applied to other dynamics than Eq (3), as well as other definitions of sensation relatively to these dynamics.

**Color sensation and not perceived color.** In our attempt to build the most general mathematical framework defining color matching experiments, test and comparison images $J^{test}, J^{comp} \in \mathfrak{C}^{\mathbf{R}}$ are *not* supposed to have a simple or identical geometry. Formally, $J^{comp} = J^{comp}[c]$ is an image parameterized by some color $c$, which the observer can adjust in the experiment, until reaching a perceptual match for the value $c = c^{match}$ between $J^{test}$ and $J^{comp}$ at the points of interest $x^{test}$ and $x^{comp}$. During the search process, the observer explores the family of possible comparison images $\{J^{comp}[c]\}_{c \in \mathfrak{C}}$. Matching is relative to the specific pair of points $(x^{test}, x^{comp})$, because there is no reason for the final comparison image $J^{comp}[c^{comp}]$ to be *everywhere* perceptually equivalent to the test image (see Fig 2).

We believe that it probably *does not make sense* to define "*the*" perceived color, which would be an element of $\mathfrak{C}$. As an illustration of this difficulty, one would say that "the" perceived color in $J^{test}$ of Fig 2 is some green $c^{match}$. But if the comparison surround had been replaced by a lighter or darker gray, the resulting matching colors would be different from the previous one, for they depend on the comparison background. Yet, the perception of a test color *should not depend* on the comparison image. Instead, we find more appropriate to talk about a *color sensation* elicited by an image $J$ at some point $x$.

**Definition (color sensation).** Let $J$ be a fixed image inducing the cortical image $I(r) \coloneqq J(x)$ with $r = \chi(x)$ where $\chi$ is the retinotopic mapping. Suppose that there exists a unique stationary solution to which the dynamics of Eq (3) converges, denoted $a_J(\cdot, \cdot, \infty) \coloneqq \lim_{t \to \infty} a_J(\cdot, \cdot, t)$. Then the *color sensation* generated by $J$ and perceived at a cortical point $r_0$ is the following element of $\mathbb{L}^\infty(\mathfrak{C}_{opp})$:

$$a_J(r_0, \cdot, \infty) : \mathfrak{C}_{opp} \to [0, 1].$$

Under some conditions, there exists a unique stationary state, which is linearly stable (see S3 Appendix). It is however possible that there exist several stationary solutions: in this case, we should use a more sophisticated definition of color sensation, which is outside the scope of this paper. Color sensation at $r_0$ is thus a function on $\mathfrak{C}_{opp}$ which represents *how the hypercolumn responds to any color*. This concept can be extended to a group of hypercolumns by considering a collection of color sensations. Our representation is much richer than defining "the" perceived color, since we have instead a whole function attached to each hypercolumn, as an echo of the complexity of color perception with spatial context.

**Color matching as a mathematical projection.** We propose that matching can be mathematically described as a projection. This process corresponds to *matching two brain states*, whatever the perception being matched, such as color, texture, touch, pitch, timbre, or any other feeling. Such a formal description implies that color sensations can be compared in a quite objective manner though matching experiments; and more generally, perception is likely to be a *relative concept* that can be assessed through comparisons as objective as possible, instead of an absolute one.

In the following, the cortical points where matching takes place are denoted by $r^{test}$ and $r^{comp}$, and the corresponding color sensations elicited by $f^{test}$ and $f^{comp}[c]$ are denoted by $a^{test}$ and $a^{comp}[c]$ respectively.

**Proposition (matching as a projection).** A color matching experiment consists in choosing $c^{match} \in \mathfrak{C}$ so that $a^{comp}[c^{match}]$ is *closest* to $a^{test}$, *i.e.*,

$$c^{match} := \arg\min_c \ \mathrm{dist}(a^{test}, a^{comp}[c]) \tag{9}$$

where dist is a perceptual similarity criterion.

For example, one can take $\mathrm{dist}(a^{test}, a^{comp}[c]) = \|a^{test} - a^{comp}[c]\|_{\mathbb{L}^{\infty}(\mathfrak{C}_{opp})}$. In S3 Appendix, we show that under appropriate conditions, we can smoothly parameterize color sensations $a^{comp}[c]$ with respect to $c$. Hence, color matching is formally the projection of $a^{test}$ onto a nonlinear manifold whose elements are $\{a^{comp}[c]\}_{c \in \mathfrak{C}}$.

## Data

**Reference data.** We used the data in the figures of [18] and [17]. Usage of the data was kindly approved by the authors. In order to apply our model to this dataset, we suppose that color-tuned cells in the hypercolumns are of three types, each tuned to one particular 1D color axis, and that color matching is made independently on each axis. Because the data mainly involved $s$ chromaticity, we applied the model only along the $s$ axis independently of $l$ and $Y$ axes.

**Personal data.** Throughout the experiments, we used two particular test and comparison backgrounds filled with Yellow ($HSL = (60°, 50\%, 50\%)$) and Gray ($HSL = (0°, 0\%, 50\%)$), respectively. A.S. performed color matching experiments as in Fig 3 (in a spirit similar to [3] who worked with achromatic colors). Experiments used a standard computer screen, in a dark room. Squares were presented against a black background. Asymmetric color matching was repeated for several colors tested with the Yellow surround, at regularly spaced locations in the three-dimensional HSL color space. We obtained a *vector field of color shifts* in the HSL space associated to the pairs ($c^{test}$, $c^{match}$), as seen in Fig 5 (note that representing shifts as a vector field is not new [9]). Our final data was obtained by averaging the shifts along the Luminance coordinate, and lived in the chromatic disk only (see the Results section), as we are mainly interested in chromatic shifts. The treatment of luminance seems to be more complex, and has to be separate from that of chromaticity.

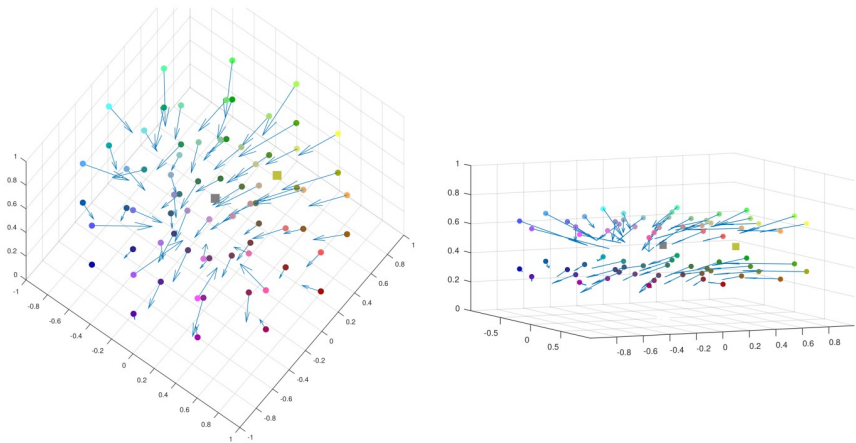

**Fig 5. Yellow pushes towards blue.** Using fixed Yellow test and Gray comparison surrounds (symbolized as squares in the figure) similarly to Fig 3, we obtain a vector field of shifts ($c^{test}$, $c^{match}$) in the HSL space (symbolized as spheres and arrows). There is clearly a "yellow pushes towards blue" phenomenon, where contrast wins.

https://doi.org/10.1371/journal.pcbi.1007050.g005

To apply our model to this dataset of *2D* shifts, we suppose that color-tuned cells in the hypercolumns are of two types, one tuned to *2D* HSL chromaticity, and the other to *1D* Luminance. As before, color matching is supposed to take place independently within each subspace, and we only considered what happened in the chromatic disk.

## Results

To validate our model and compare it to experimental data, we regressed the scalar parameters

$$q := (\mu_c, v_c, \alpha_c, \beta_c, \mu, v, \alpha, \beta, \mu_h, \sigma_h, \gamma)$$

involved in the Gaussians and the activation function (see Eqs (4), (5), (7) and (8)), to the psychophysical data described above. The regression consists in minimizing some energy $E(q)$ that is the sum of the squared errors between the predicted matching color $c_q^{pred}$ and the actual matching color $c^{match}$ found in the experiments. Details of this difficult numerical task are provided in S2 Appendix, where we state three algorithms for reproducing the Color Neural Field dynamics, simulating a color matching experiment, and performing the regression. The latter two algorithms can be used for *fitting general neural fields to general matching data*. Here is a summary of our results.

- We illustrate the role of the connectivity kernel $\omega$ and of the cortical input $H$ (Figs 6 and 7).

- We simulate the dynamics of Eq (3) in the case of a purple/lime patterned cortical image $I$ shown in Fig B in S2 Appendix (Fig 8).

- Our model is *able to predict the data* from observers 'MC' and 'AZ' of [18] respectively, after regression (Fig 9).

- The regressed functions $f$ and $g$ corresponding to observer 'MC' are displayed, and we illustrate the concept of color sensation (Figs 10 and 11).

- We regress the model to the data of [17] and *succeed in explaining the nonlinear behavior of the shifts* along the $s$ chromaticity of the test color (Fig 12). This constitutes an important result in the present work, since it initially motivated the latter.

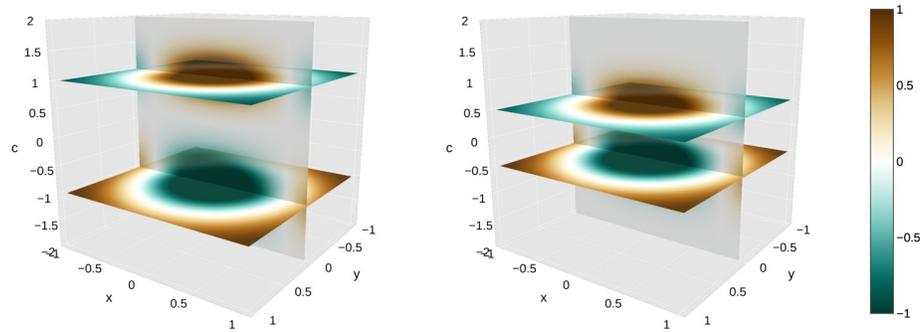

**Fig 6. Connectivity kernel in 3D physical and color space.** Left. $W_1$ defined with $r_0 = 0$ and $c_0 = 1$. Right. $W_2$ defined with $r_0 = 0$ and $c_0 = .5$. The color bar extends between $-1$ and $1$, going from dark green (negative values, strong inhibition) to dark orange (positive values, strong excitation). *We provide an interactive 3D animation of the connectivity kernel* $\omega(r_0, c_0, \cdot, \cdot)$ *for all values of* $c_0 \in \mathfrak{C}_{opp}$ *in* S1 File. *For varying* $r_0$, *the kernel is just spatially translated along* $r_0$. *However, for varying* $c_0$, *the positive and negative gaussian kernels in f follow* $c_0$ *and* $-c_0$, *collide when* $c_0$ *goes through zero, then exchange of position when* $|c_0|$ *grows again.*

https://doi.org/10.1371/journal.pcbi.1007050.g006

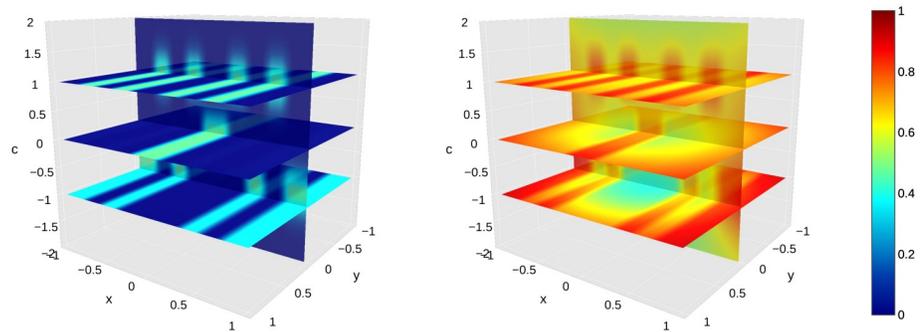

**Fig 7. Cortical input and color sensations in 3D physical and color space.** Left. Cortical input $H$. Right. Color sensations $a_\infty$. In both subfigures, the color bar extends between 0 and 1, and is set so that small variations are easily seen. *We provide an interactive 3D animation of the evolving activities* $a(\cdot, \cdot, t)$ *along the iterations of the fixed point algorithm in* S2 File. *The convergence is quite fast and 15 iterations are sufficient.*

https://doi.org/10.1371/journal.pcbi.1007050.g007

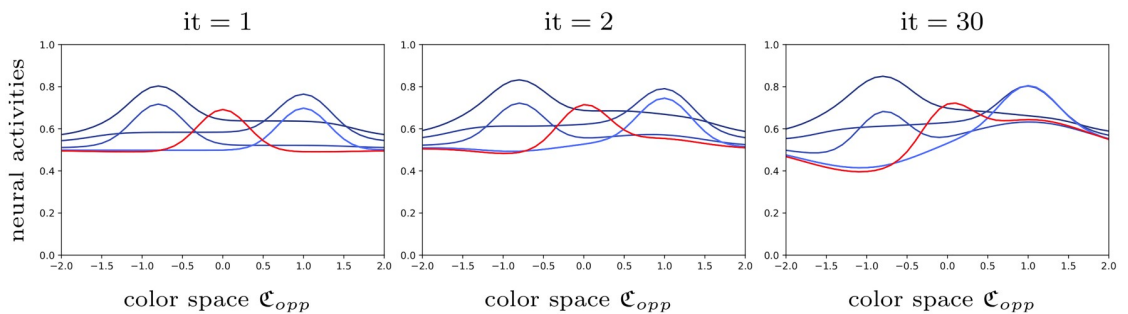

**Fig 8. Dynamics of the color neural field** Eq (3)**.** The neural activities are plotted after a. one, b. two, c. thirty iterations, and convergence is reached after fifteen iterations. The red curve indicates the activity $a_q^{sen}(r_0, \cdot)$ of hypercolumn $r_0$ corresponding to the test stripe. Other blue curves correspond to spatial points $r_i$ located on surrounding stripes. Notice that only four and not eight different curves are seen, because of the axial symmetry artificially introduced to facilitate numerical computations, as explained in S2 Appendix. *A video of the dynamics is provided in* S1 Video.

https://doi.org/10.1371/journal.pcbi.1007050.g008

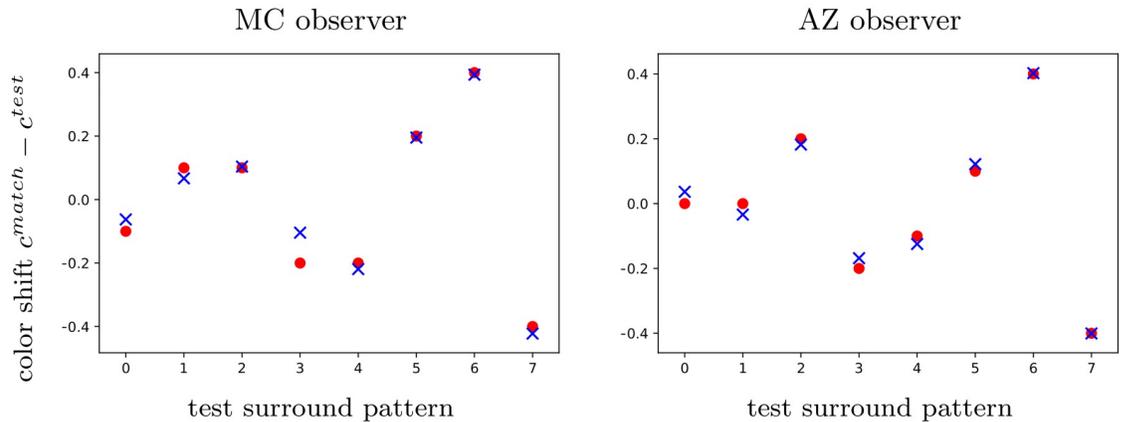

**Fig 9. Our tuned model predicts color shifts from [18].** Left. Observer 'MC'. Right. Observer 'AZ'. Red dots indicate experimental data, while blue crosses stand for predicted matching comparison colors. The data has been averaged over three sets of experiments, as detailed in the original article. The ordinate corresponds to the color shift, expressed in $\mathfrak{C}_{opp}$ coordinates with $c = s - 1$. The abscissa $i = 0, \ldots, 7$ refers to the test pattern: p/p, l/l, p/w, l/w, w/p, w/l, p/l, l/p, where p stands for purple, l for lime, w for white.

https://doi.org/10.1371/journal.pcbi.1007050.g009

- Even when the model is regressed to observer 'AZ' [18] in a setting where the $s$ chromaticity is not changed, it is still able to predict a nonlinear behavior similar to that observed when $s$ is changed such as in [17] (Fig 13).

- We are also *able to explain the vector field of shifts* ("yellow pushes towards blue" phenomenon of Fig 5) in the chromatic disk, for our data (Fig 14).

Before going further, let us point out some important facts.

1. We do not claim that the regressed parameter value is indeed a global minimizer of the energy $E(q)$ (see S2 Appendix), but only that it is sufficient for the algorithms to approximately mimic the experimental data in a satisfying way.

2. For each regressed parameter, and for one typical image input (purple/lime pattern), we empirically checked that the activity $a_\infty$, to which our Algorithm 1 with $dt = 1$ (S2 Appendix) converges, is indeed a stable stationary solution of Eq (3) to which any solution converges in time. Furthermore, we numerically checked that this stable steady state is unique, so that we can call its restriction to the hypercolumn $r_0$ a color sensation.

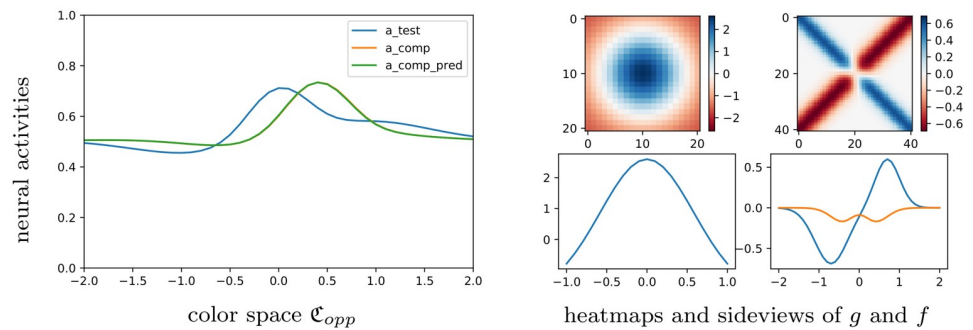

**Fig 10. Comparing color sensations for real and predicted matching color.** Left. Color sensations $a_q^{test}$, $a_q^{comp}[c^{match}]$ and $a_q^{comp}[c^{pred}]$. Right. Functions $f$ and $g$ for the regressed parameter value $q = q_{MC}$. Right, up. Heatmaps for $g$ and $f$. Right, down. Corresponding side views.

https://doi.org/10.1371/journal.pcbi.1007050.g010

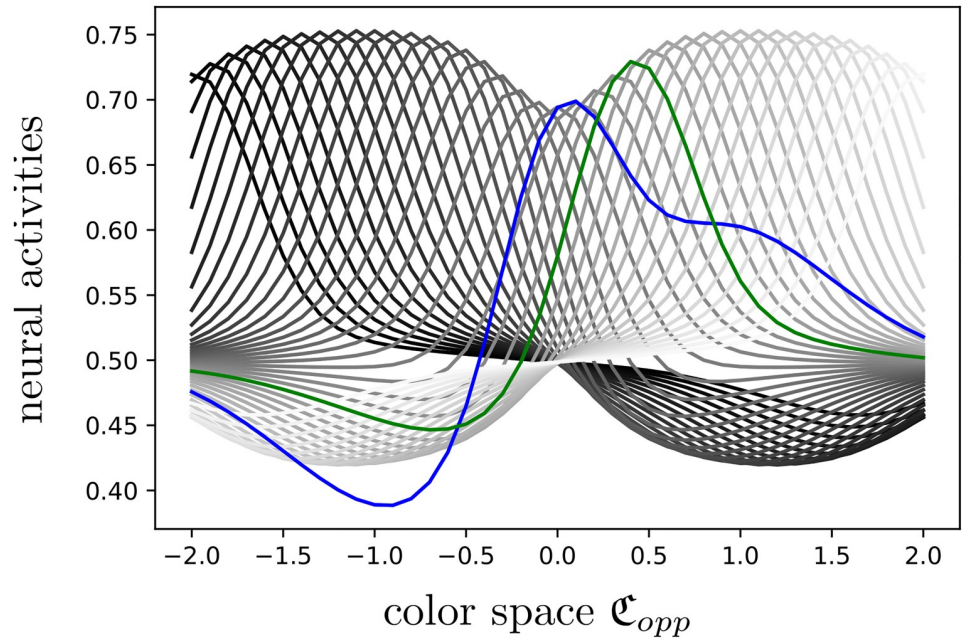

**Fig 11. Color matching is a projection.** Blue and green curves are as in Fig 10 and correspond to $a_q^{test}$ and $a_q^{comp}[c^{match}]$. After regression, the green one should be the nearest to the blue one with respect to the $\mathbb{L}^\infty$ norm, among all the curves $\{a_q^{comp}[c]\}_c$ generated by the family of comparison images $\{I^{comp}[c]\}_{c \in \mathfrak{C}_{opp}}$ (shown in gray).

3. Before undertaking any solid interpretation of the regressed parameter values, edge effects due to the finite size of the spatial domains in the numerical implementations still have to be carefully considered, which is outside the scope of this article.

Note that for all figures except Fig 14, we consider the settings of [18] and [17], in which the color space is one-dimensional (see Materials and methods section). In Figs 6 and 7, the connectivity values, the cortical input $H$ and the neural activities $a_\infty$ are illustrated as 3D heatmaps defined on $\Omega \times \mathfrak{C}_{opp}$. The 3D space is the product of the 2D physical space by the 1D color space, parameterized by $x$, $y$ and $c$ respectively. In Figs 6, 7, 8, 10 and 11, the dynamics, inputs and connectivity values are parameterized by the value $q = q_{MC}$ optimized on observer 'MC' of [18] (see Fig 9).

## Connectivity kernel and cortical input

In Fig 6, we set $W(\cdot, \cdot) := \omega(r_0, c_0, \cdot, \cdot)$, and display it for two values of $(r_0, c_0)$. The two configurations of Table 2 (excitatory or inhibitory connections) can be clearly seen in the figure. In fact, most of the neural masses have negligible influence on $(r_0, c_0)$. This occurs when $c'$ is neither close to $c_0$ nor to $-c_0$, or when $r'$ is too far from $r_0$ (not shown on the figure, because the outer variance of the DOG $g$ has a great value compared to the extent of $\Omega$). The properties of the connectivity values are analogous to double-opponent cells' center-surround behavior.

In Fig 7, we show the cortical input $H(r, c) = h(c - I(r))$, where the purple/lime patterned image $I$ is as in Fig B in S2 Appendix and corresponds to the cortical counterpart of one typical test pattern of [18]. We also show the color sensations $a_\infty$ after convergence. The input $H$ is obtained by "lifting" the image $I$ inside $\Omega \times \mathfrak{C}_{opp}$. Altitudes of maximal values hence alternate between lime ($c \simeq -1$), purple ($c \simeq 1$), and white ($c \simeq 0$). The shape of $H$ heavily determines

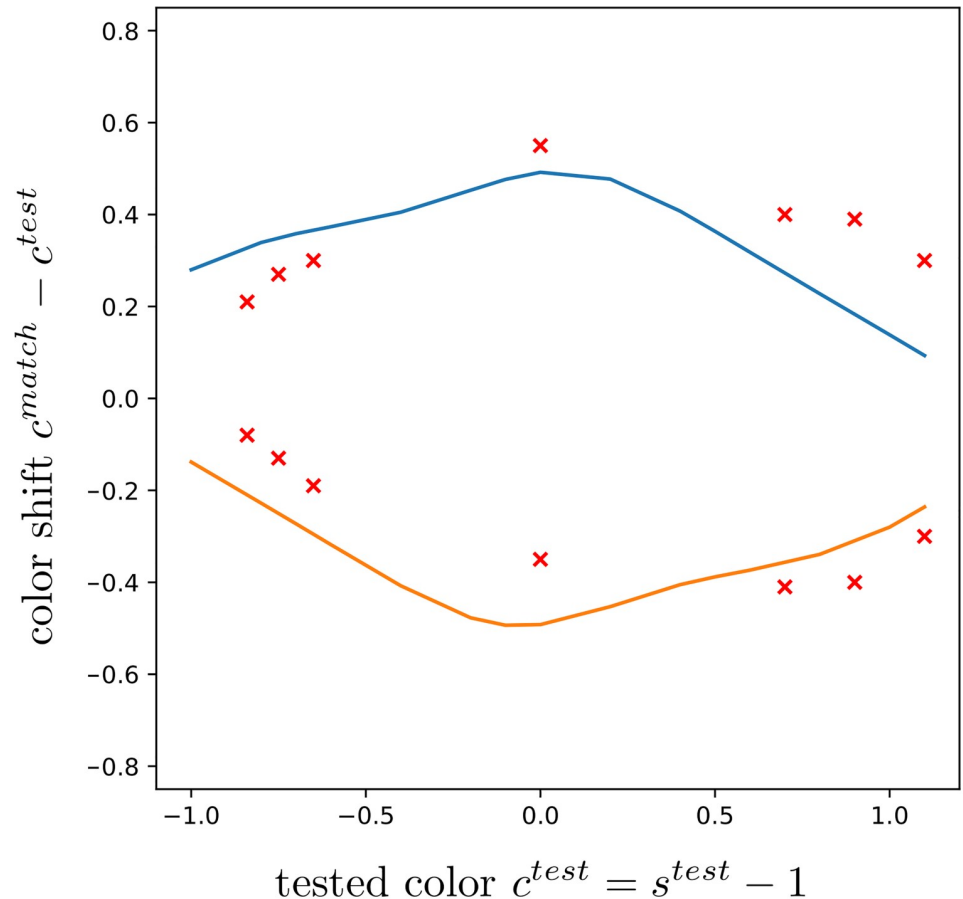

**Fig 12. Our tuned model is able to reproduce nonlinear data from [17] Fig 3.** The blue and orange curves correspond to our predicted shifts along the tested chromaticity given by the abscissa. The data above and below the zero line correspond to a fixed purple/lime or lime/purple pattern, respectively. Red crosses indicate the means of shifts across four subjects for seven tested values, and stand as groundtruth (refer to [17] for details).

https://doi.org/10.1371/journal.pcbi.1007050.g012

that of the final activities $a_\infty$, since it has the role of cortical input. It can be noticed that $a_\infty$ reaches values lower than 1/2 in the bottom part of the heatmap.

## Simulation and regression results

In order to emulate a color matching experiment, the first building block of our algorithm is to simulate the Color Neural Field dynamics in Eq (3), as shown in Fig 8, and accordingly to Algo 1 in S2 Appendix (with $dt = 1$). Once again, the cortical image is given by Fig B in S2 Appendix.

Color matching can then be emulated by applying Algo 2 in S2 Appendix. The parameters have to be regressed to the experimental data in order to reproduce the color shifts. Fig 9 shows that our model is able to explain the shifts measured for the observers called 'MC' and 'AZ' in [18], by using the regressed values $q_{MC}$ = (0.60, 0.69, 0.30, 0.40, 4.42, 1.82, 0.58, 8.35, 0.47, 0.30, 1.80) and $q_{AZ}$ = (0.60, 0.69, 0.31, 0.40, 4.42, 1.81, 0.60, 8.35, 0.47, 0.30, 1.80), respectively. The slight difference observed between the parameter values could partly account for subject differences.

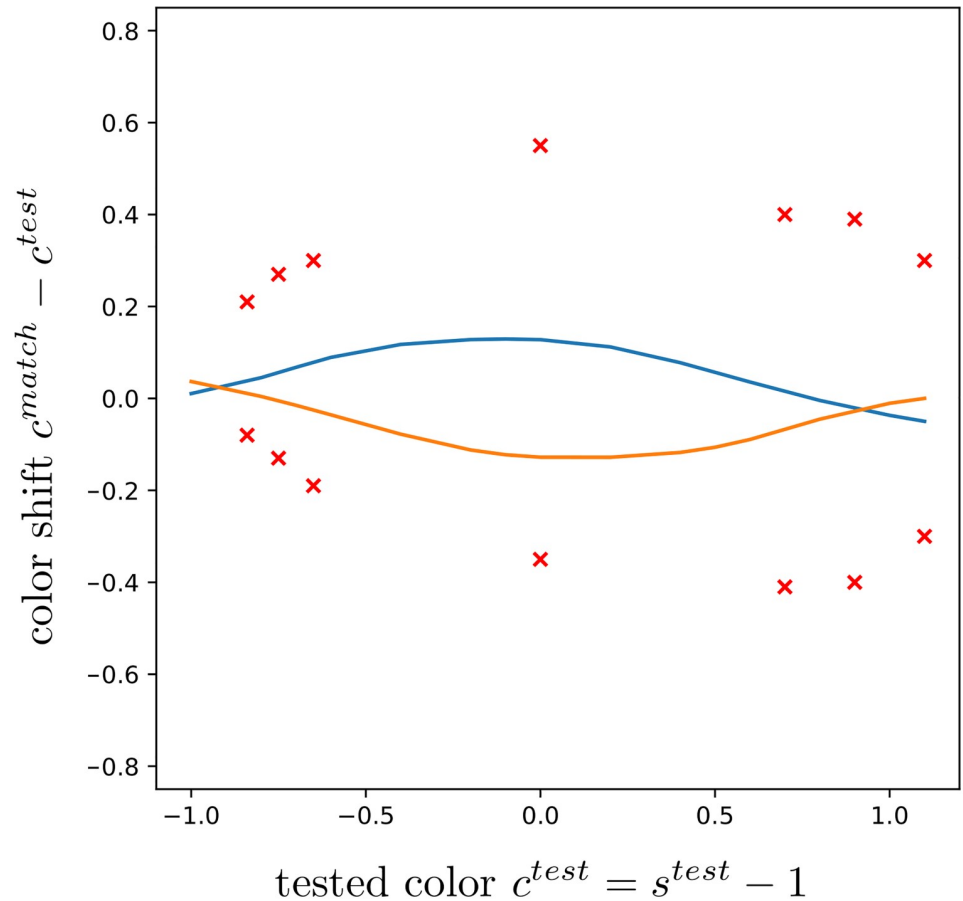

**Fig 13. Our model reproduces nonlinearity even when tuned to other data.** We use the parameter value $q_{AZ}$ corresponding to Fig 9 (right) while emulating the experiments of Fig 12.

https://doi.org/10.1371/journal.pcbi.1007050.g013

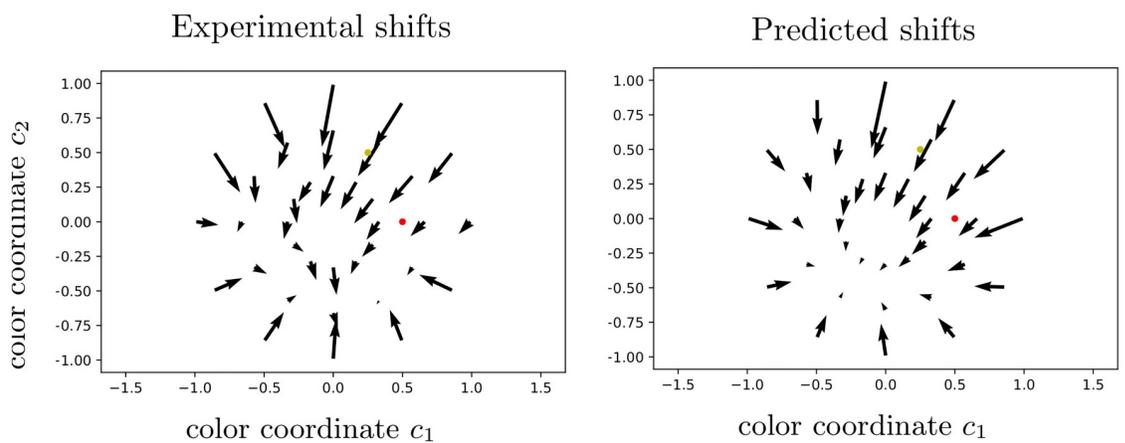

**Fig 14. The color neural field model reproduces nonlinear shifts in the chromatic disk.** Left. 36 pairs of experimental data points (test and matching colors) in the HSL chromatic disk at constant luminance, resulting from averaging the shifts (refer to Data section and Fig 5). Right. Predicted results.

https://doi.org/10.1371/journal.pcbi.1007050.g014

We compare in Fig 10 (left) the color sensations $a_q^{test}$, $a_q^{comp}[c^{match}]$, and $a_q^{comp}[c_q^{pred}]$ (refer to S2 Appendix for the notations), taken at the point of interest $r_0 = (0, 0)$, and with $q = q_{MC}$. They are generated by the purple/lime test image (as before), the comparison image filled with the experimental value $c^{match}$, or with the predicted matching value $c_q^{pred}$, respectively. After regression, $c_q^{pred}$ becomes close to the experimental value $c^{match}$, so that the two corresponding curves nearly coincide.

Among all curves $\{a_q^{comp}[c]\}_c$, $a_q^{comp}[c^{match}]$ should be the nearest one to $a^{test}$ with respect to the $\mathbb{L}^\infty$ norm, and we illustrate this in Fig 11. The qualitative difference between the test and comparison curves mainly comes from the difference of complexity of their respective inputs: the test image has a complicated pattern, while the comparison images are simpler.

We confront our model to the nonlinear shifts observed by [17] in Fig 12. We find that it is able to reproduce the data after regression, with the fitting parameter value $q_{nonlin}$ = (0.42, 0.71, 0.63, 1.16, 4.43, 1.72, 0.56, 6.35, 0.47, 0.30, 1.80). This non-trivial result, alongside the results in Fig 9, provides a strong justification to our framework. In Fig 13, instead of $q_{nonlin}$, we used the value $q_{AZ}$ that explained the data on Fig 9 (right) observed by [18]. Let us recall that in the experimental settings of [18] and Fig 9, $s^{test}$ was fixed while the surround varied, unlike in the settings of [17] and Fig 12, where conversely for some fixed test patterns $s^{test}$ was changed. The predictions in Fig 13 are hence not close to the ground truth data, especially at the endpoints where the orange and blue curves are crossing. Shifts are also smaller in magnitude. However, it is remarkable that the model predicts a similar trend with respect to $s^{test}$. Crossings are also quite expected at $s^{test}$ of magnitude great enough, for which a reversing of the shift direction is plausible. For too great magnitudes, the shifts are likely to become negligible. We obtain similar results by using $q_{MC}$.

Finally, as a further important validation, we show in Fig 14 that our framework is also efficient in explaining the vector field of shifts in the chromatic disk (Fig 5). The convergence towards the opposite blue is made obvious, with the regressed parameter value $q_{HSL}$ = (0.73, 0.15, 0.52, 0.68, 4.41, 1.84, 0.51, 8.35, 0.47, 0.30, 1.80). As a remark, the sampling resolution of the 2D color space is quite low for computational reasons, so that the meaningfulness of parameter values has to be carefully considered. We also obtain a smoother result than in the experimental data, as a SoftMin method is used to search for optimally matching colors (see associated code).

## Discussion

*This work constitutes a first attempt at building a neural field framework for color perception which explains psychophysical data in a consistent manner. Our model addresses the important problem of color-space interactions (see* Introduction*), while unifying chromatic contrast and assimilation.* It relies on the hypothesis that color-tuned cells in layers 2/3 of V1 are physiologically structured in hypercolumns, sharing spatially similar receptive field positions and tuned to a continuum of colors. We have tested this model on psychophysical data from two kinds of experimental settings (that of Monnier and Shevell [17, 18] and ours), by supposing that cells separately encode three 1D color axes or a 2D chromatic plane alongside a 1D achromatic axis, respectively, and that color matching takes place independently within each subspace (see Data section). These two assumptions, although very different, have both led to promising results.

Our model provides a Color Neural Field as well as a color matching framework, which is in fact suitable to general dynamics and independent of the particular equations we have used in this work (3) (see S2 Appendix). The notion of color sensation connects these cortical and perceptual levels, and is *a nonlinear object extracted as a percept from an assembly of neurons.*

It is compatible with the nonlinearity of the responses of V1 neurons with respect to cone signals, which has been shown to constitute a distinctive feature compared to neurons in the LGN and the retina [2, 23]. Furthermore, color sensation takes into account a whole distribution of activities, in contrast to single-neuron concepts. This idea had already been suggested in the context of *population coding* [65–67], but with different perspectives (where preferred stimuli are integrated over several neurons to deduce the preferred stimulus of the whole population, while here dynamic activities elicited by one stimulus are spatially integrated to produce sensation, which still contains information about each neuron). Finally, note that considering matching as a projection is an idea quite close to that of inversion in the corresponding-pair procedure proposed in [68], which predicts colors that produce the same sensation in dichromat and trichromat viewers. This concept of inversion corresponds to the particular case when the brain state or percept (such as color sensation in our sense) is in fact at zero distance from the set on which it is projected (which in our case does not happen in practice).

## Distinction from color constancy problems

The present work has to be distinguished from those dealing with color constancy problems. *Color constancy* is the ability of humans to guess the reflectance of objects despite very different illumination conditions [69]. Reflectance is a property characterizing matter, defined as the proportion of luminous energy reflected by its surface. In Eq (1), it is linked to the spectral power distribution $\mathcal{P}$ of the incident illuminant (daylight, lamp) through the relation

$$\mathcal{C}^x(\lambda) := \mathcal{P}^x(\lambda)\mathcal{R}^x(\lambda),$$

where reflectance and illuminant power are taken at the point of the scene which sends light to $x$ in the retina. The phenomenon of color constancy, first studied by E. Land who proposed the Retinex algorithm [70, 71], has since been the subject of much research. Given cone inputs $(L, M, S)$, how does the brain retrieve the spectral reflectance $\mathcal{R}$ of an object? By contrast, we are not interested in how color is linked to reflectance, but rather in how identical stimulation of the cones can lead to different sensations because of spatial context.

## Three meanings of "color"

In view of the previous discussion, to the naive question "What is the color of this object?", we see that at least three types of answers are theoretically possible. First, identify the colored material likely to produce such a visual effect, *i.e.*, guess its reflectance $\mathcal{R}$; second, report the color $[\mathcal{C}]$ produced by cone stimulation; and third, describe the color sensation, as introduced in this work. In practice, none of these tasks is trivial, even regarding color naming, since thought and language related to color involve complex cognition [72].

Summarizing the previous ideas, the three punctual relationships

$$\mathcal{R}_1^{x_1} \quad = \mathcal{R}_2^{x_2}$$

$$[\mathcal{C}_1^{x_1}] \quad = [\mathcal{C}_2^{x_2}]$$

$$\mathfrak{S}_1^{x_1} \quad = \mathfrak{S}_2^{x_2}$$

are independent, because there are situations where any of them can hold or not. The reader can be convinced by simple examples. For instance, the same object seen in daylight or under shadow has different colors because $[\mathcal{C}_{daylight}] \neq [\mathcal{C}_{shadow}]$, but we can easily recognize it and guess that $\mathcal{R} = \mathcal{R}_{daylight} = \mathcal{R}_{shadow}$. On the reverse, two objects having the same color can be guessed as being made of different materials, if their surroundings are not the same, that is, $[\mathcal{C}_1] = [\mathcal{C}_2]$ but $\mathcal{R}_1 \neq \mathcal{R}_2$. Another example is given by color induction, as largely discussed here. Because of the surrounding context, two colors $[\mathcal{C}_1] = [\mathcal{C}_2]$ deriving from identical L,M,S cones stimulation can be *perceived* as different color sensations $\mathfrak{S}_1 \neq \mathfrak{S}_2$. The reverse situation can occur, where different colors are perceived as similar with different contexts.

However, if the whole retinal space is taken into account, then the sets

$$\{\mathcal{R}^x\}_{x \in R} \tag{10}$$

$$\{[\mathcal{C}^x]\}_{x \in R} \tag{11}$$

$$\{\mathfrak{S}^x\}_{x \in R} \tag{12}$$

are linked to one another. While papers on color constancy are devoted to the link between Eqs (10) and (11) [69, 73], ours focuses on the relationship between Eqs (11) and (12). Our framework supposes that $\{\mathfrak{S}^x\}_{x \in R}$ depends on the sole knowledge of $\{[\mathcal{C}^x]\}_{x \in R}$.

## Our model is not an image processing model

Our model should **not** be considered as an image processing algorithm. First, our work involves a psychophysically and physiologically relevant neural structure, while image processing models often lie at the conceptual level of an image. In the literature, many algorithms are designed to make the image perceptually better (contrast enhancement, histogram equalization, *etc.*) and allow for image compression, in accordance with various perceptual criteria, an early work being [74] for example. Some of them are computational models simulating perception, and inspired by neural mechanisms underlying vision, as in [75], and more recently [76]. Taking an image as input, their algorithms produce another image as output, which represents "the" perceived color image, in order to reproduce color induction effects. In fact, the simple convolutional receptive-field model of [18] is already a basic version in this family of algorithms. In the more sophisticated approach of [75], the image evolves according to a Wilson-Cowan dynamic, which is akin to descending the gradient of some energy, leading to histogram equalization. M. Bertalmío also proposes a method to reproduce the lightness matching data of [77], in particular some assimilation or contrast effects (in the usual global sense, not at the local scale). His attempt to match psychophysical data based on neural-like dynamics can be considered similar to ours. However, shifts are estimated through a direct algebraic computation of the values found at different locations of the disks after convergence. This prevents any straightforward generalization of the method to spatially more complicated patterns, an impediment which we already mentioned in the Introduction for the model of [18].

More importantly, *a major conceptual problem in this image-based approach is that, it seems inappropriate to simulate perception by producing an output image of the same nature as the original one, that represents "the" perceived color image*. Indeed, the original and final images *do look different*: the final image hardly represents our perception of the original one, since it is also processed by the brain. This ambiguity should be clearly discussed when dealing with the output image. For instance, [74] proposes to map color images into an opponent perceptual space, where they are processed and compared through a perceptual metric, that has a role clearly different from the usual RGB color space of images.

## The need for a universal model

It is the fate of any model to meet its own limitations. Below, we depict non-exhaustively some questionable features or drawbacks encountered by ours.

**Physiological interpretation of the model.** In the Results section, we suggested that the shape of the connectivity kernel might be related to the behavior of a double-opponent cell, and that the cortical input to $(r, c)$ might be relayed by single-opponent cells in the LGN, or in layers $(4C\beta)$ and $(2/3, 4A)$ of V1. The double-opponent cells of our framework would then be connected *over* single-opponent cells (similarly to the color-orientation model of [42]). These analogies should *not* be regarded as firm assertions, but rather as possible links between our model and current knowledge about the physiology of V1 neurons. At least, double-opponent cells are suspected to play a role in simultaneous contrast, and single-opponent ones in assimilation (see Introduction). If pursuing further our analogy, both populations would be in fact involved in these effects, and a combined action of the two is indeed expected [2, 23, 78]. However, our model has two important drawbacks regarding its biological interpretation. First, the values found for the inner ON-center diameter of the DOG $g$, after the different regressions to data of [17, 18] (see Results), approximately correspond to 2° of visual field and cover about 7 stripes (the test one, and three on each side). This value is likely to be two to six times too large, because one would rather expect a frequency selectivity in the range of 1–3 cyc/deg [79]. It is probable that we did not find more plausible values for the parameters in $g$ because of the small spatial domain we used in numerical computations. Yet, the numerical regression is quite complex (minimizing with respect to shifts, which themselves involve the minimization of matching) and has to satisfy numerous criteria (closeness to data, nonlinearity, non-saturation of the activity, fast convergence in time of the dynamics, and reasonable duration of the regression). Looking for more biologically plausible values, using a larger spatial domain and finer discretization steps, was hence out of reach in reasonable time for the current work. Furthermore, another defect of the model is that it only considers concentric center-surround receptive fields for double-opponent cells, as in early works [35, 36], although anisotropic receptive fields have been found later [2, 23, 78], which account for their sensitivity to orientation.

**Color-orientation and color-edge interactions.** We have not explored the link between orientation and color, contrarily to the complementary work of [42] (see Introduction). Instead, we have addressed color-space interactions, and compared the model to (spatially non-trivial) data. It is worth noting that, cells are supposed here to be tuned to colors in one to three dimensions, whereas in their model, cells were tuned to different hues (organized along the circle), with luminance and saturation being treated as input and amplitude of the activity, respectively. Also, they introduce a two-color external signal, which allows two different hues to have effect on the same hypercolumn, as they suppose is the case when the brain expects to see one color but receives another stimulus instead. This requires however to modify the input involved in the dynamics. Unlike them, we consistently consider one color input per location, and define color sensation in a flexible manner which can in particular account for a "two-color" impression.

Orientation and edges should both be integrated into the construction of a more universal color-space model. Edges are thought to have an important role due to "visual edge contrast" [80] in the perception of a colored surface, and to be selectively processed in V1 [79]. In particular, they may contribute to watercolor effects [81], in which a local feature (two thin borders of dissimilar colors) spreads into a global impression (one of the delimited area seems to be faintly colored by the inner border). A simple integration over the spatial domain, as made here, cannot sufficiently account for the effect of a colored edge over the entire perception.

**Further remarks.** We used an ideal mathematical setting to ease the analysis, but it is quite improbable that an opponent representation of $\mathfrak{C}$ enjoys perfect symmetry. Moreover the geometry and structure of $\mathfrak{C}$ is still not entirely understood. It seems reasonable however to talk about a nearly symmetric space, and work in the biggest symmetrically opposable space we can define inside $\mathfrak{C}$.

Furthermore, we did not compare our model to data varying along the luminance axis, because it certainly deserves a special treatment, separately from the chromatic plane. An easy adaptation of our model would be to consider anisotropic gaussian kernels in $f$ and $g$, or use a non-separable connectivity kernel $\omega$, which could be a finite sum of separable kernels $\omega_i = f_i \otimes g_i$. This should already allow different treatments of the luminance and chromatic dimensions.

Finally, we insisted on using color sensations (functions on $\mathfrak{C}$), and not perceived colors (an element in $\mathfrak{C}$), to describe the perception of a color. This would indeed lead to confusions (input *vs.* perceived images), and incompatibility with the matching experiments (see Discussions above and the Materials and methods section). However, in our framework it is still possible to define "the" perceived color as the one $c_{max} \in \mathfrak{C}$ which maximizes the neural activity in hypercolumn $r_0$, similarly to the "winner-takes-all" law of orientation perception. The problem of this approach is that our model would then predict positive afterimages, given that the input image is positively added in the dynamics.

## Towards color hallucinations

Just as for neural field models modeling orientation vision, we can study bifurcations of the solutions of Eq (3) around stationary states [49, 53, 56, 82]. Under some hypotheses of symmetry and periodicity, we can predict, using equivariant bifurcation theory, the emergence of visual patterns or "planforms". In the same fashion as [49] who explained orientation-based geometric hallucinations, a color neural field model can predict patterned color hallucinations. Future psychophysical experiments, may confirm this and support the relevance of this kind of model for color vision.

## Conclusion

Our work addresses the question of color-space interactions, by providing a color neural field model alongside a general framework to account for matching experiments. We propose to consider color matching as a mathematical projection, in agreement with the principles of psychophysics, where subjective notions are assessed by means of objective procedures. Our neural field unifies assimilation and contrast at the cortical level, and relies on the idea of color opponency. The notion of color sensation that we introduce bridges the gap between these cortical and perceptual levels, and is a nonlinear percept involving a whole distribution of neurons.

This framework allows the study of psychophysical phenomena such as color induction, by taking advantage of a classical computational neuroscience tool. To our knowledge, this is the first color neural field model consistent with psychophysical data and compatible with physiological findings. The assumption that V1 is organized into a structure similar to color hypercolumns has still to be experimentally proved though. We believe that the proposed framework could possibly be adapted to other perceptual situations, such as hearing or touch.

## Supporting information

**S1 File. 3D animation of connectivity kernel.**
(HTML)

**S2 File. 3D animation of evolving neural activities.**
(HTML)

**S1 Video. Evolution of neural activities at different locations in the cortex.**
(AVI)

**S1 Appendix. Mathematical definition and properties of color space.** We rigorously define the color space $\mathfrak{C}$ and its classical representations; we explain the link between color matching functions and cone sensitivities.
(PDF)

**S2 Appendix. Simulations and regression.** We detail the numerical implementations of our model. We provide three algorithms for the Color Neural Field equations, color matching, and regression of the model to psychophysical data.
(PDF)

**S3 Appendix. Further information about the model.** We detail some properties of the dynamics of the neural field; we describe the set on which color matching is a projection.
(PDF)

## Acknowledgments

The authors thank Patrick Monnier and Steven K. Shevell for providing them their data gracefully. A.S. acknowledges the fruitful discussions she had with Daniel Bennequin. The authors are grateful to Grégory Faye for drawing their attention to reference [59].

## Author Contributions

**Conceptualization:** Anna Song, Olivier Faugeras.

**Data curation:** Anna Song.

**Formal analysis:** Anna Song, Olivier Faugeras, Romain Veltz.

**Funding acquisition:** Olivier Faugeras, Romain Veltz.

**Investigation:** Anna Song.

**Methodology:** Anna Song, Olivier Faugeras, Romain Veltz.

**Project administration:** Olivier Faugeras.

**Resources:** Anna Song.

**Software:** Anna Song.

**Supervision:** Olivier Faugeras, Romain Veltz.

**Validation:** Anna Song, Olivier Faugeras, Romain Veltz.

**Visualization:** Anna Song.

**Writing – original draft:** Anna Song.

**Writing – review & editing:** Anna Song, Olivier Faugeras, Romain Veltz.